\documentclass[letter]{ptptex}

\usepackage{graphicx}

\title{
Axial Vector Tetraquark with Two s-quarks
}

\author{
Yoshiko \textsc{Kanada-En'yo}$^{1}$, 
Osamu \textsc{Morimatsu}$^{2}$ and Tetsuo \textsc{Nishikawa}$^{3}$
}

\inst{
$^1$ Yukawa Institute for Theoretical Physics,
Kyoto University, Kyoto 606--8502, Japan\\
$^{2}$ Institute of Particle and Nuclear Studies, 
High Energy Accelerator Research Organization,
Ibaraki 305-0801, Japan\\
$^{3}$ Department of Physics, Tokyo Institute of Technology,
2-12-1, Oh-Okayama, Meguro, Tokyo 152-8551, Japan
}

\abst{
Possibility of an axial vector isoscalar tetraquark $ud\bar{s}\bar{s}$ 
is discussed. 
If a $f_1$ meson in the mass region $1.4-1.5$ GeV consists of 
four quarks $ns\bar{n}\bar{s}$, 
the mass of the isoscalar $ud\bar{s}\bar{s}$($\vartheta^+$-meson) 
state with $J^P=1^+$ is expected to be lower than 
that of the $f_1$ meson.
Within a flux-tube quark model, 
a possible resonant state of 
$ud\bar{s}\bar{s}(J^{P}=1^{+})$ is suggested to appear at $\sim$ 1.4 GeV 
with the width ${\cal{O}}(20\sim 50)$ MeV.
We propose that the $\vartheta^+$-meson is the good candidate for the 
tetraquark search, which would be observed in the $K^+K^+\pi^-$ decay channel.
}

\begin{document}

\maketitle


The possibility of multiquark states has been discussed for a long time
\cite{Jaffe-4q,Jaffe-h,Isgur-4q,close87,carlsonb,lipkin97,Stancu98,Sakai,Barnes,Oka-rev}.
In particularly, the possible $qq\bar{q}\bar{q}$ states have been
suggested in many theoretical efforts to understand light scalar 
mesons(for example, Refs.[\ref{Jaffe-4q},\ref{Isgur-4q},\ref{Jaffe79}]). The $4q$ states 
were proposed in the description of $f_0(600)$ and $f_0(980)$, 
where the strong attraction between 
$(qq)_{\bar{3}}$ and $(\bar{q}\bar{q})_{3}$ play an important role 
\cite{Jaffe-4q,Jaffe79}. Here, $(qq)_{\bar{3}}$ and $(\bar{q}\bar{q})_{3}$
denote the color-anti-triplet quark pair and the color-triplet anti-quark
pair, respectively.
On the other hand, the $KK$ molecule 
states were suggested to understand the properties of $f_0(980)$ and
$a_0(980)$\cite{Isgur-4q}. 
Even if the $4q$ components are dominant in a certain
meson with the conventional flavor, 
it is difficult to find an direct evidence of the $4q$ components 
due to the mixing with the conventional $q\bar{q}$ state 
via the annihilation of $q\bar{q}$ pairs.
Our main interest here is the possibility of ``tetraquark'' states 
which has the minimal 4-quark content.

The recent observation of $D_{sJ}$(2317)\cite{Ds} and 
the reports of the pentaquark baryon $\Theta^+(uudd\bar{s})$
\cite{leps}, 
revived the motivation of the 
experimental and theoretical researches on multiquarks in hadron physics,
though the existence of $\Theta^+$ yet to be well established.
One of the striking characteristics of the $\Theta^+$ is its narrow
width. For the theoretical interpretation of the narrow $\Theta^+$ state,
the possibility of the spin-parity $J^P=1/2^+$
and $J^P=3/2^-$ have been discussed by many groups 
\cite{diakonov,Oka-rev,jaffe,karliner,ENYO-penta,takeuchi,nishikawa}. 
The unnatural spin and parity is a key of the suppressed width for the 
lowest decay channel.

Now we turn to the discussion on the possibility of the tetraquarks.
If we accept the interpretation of the pentaquark as the 
$(ud)_{\bar{3}}(ud)_{\bar{3}}\bar{q}$ state
based on the diquark picture by Jaffe and Wilczek,
it is natural to expect that a tetraquark with the $ud\bar{s}\bar{s}$ content
may exist at the nearly same energy
region by replacing a $ud$-diquark with $\bar{s}$ quark. 
Firstly, one should consider the states with
unnatural spin and parity, which cannot decay into two light hadrons
(pseudo scalar mesons) in the $S$-wave channel. 
Second, the exotic color configurations 
$(qq)_{\bar{3}}(\bar{q}\bar{q})_{3}$ 
would be essential to stabilize the exotic hadrons.
Then, we propose a $J^P=1^+$ $ud\bar{s}\bar{s}$ state 
with the $(qq)_{\bar{3}}(\bar{q}\bar{q})_{3}$ configuration 
as the candidate of narrow tetraquark states. 
It should be stressed that two-body $KK$ decays from any 
$J^P=1^+$ $ud\bar{s}\bar{s}$ state are forbidden because of the
conservation of the total spin and parity. The lowest threshold energy 
of the allowed two-body decays is 1.39 GeV for the $KK^*(895)$ channel.
If the mass of the $J^P=1^+$ $ud\bar{s}\bar{s}$ state
lies below(closely to) the $KK^*$, two-meson decay channels are (almost) 
closed, and hence, its width must be narrow. 

The tetraquark $ud\bar{s}\bar{s}$ states were 
discussed in Ref.[\ref{Jaffe-4q}], and noted as $E_{(KK)}$-mesons.
In the MIT bag model\cite{Jaffe-4q},
the theoretical mass for the isoscalar $ud\bar{s}\bar{s}(J^P=1^+)$ 
state was predicted to be 1.65 GeV.
Recently the isoscalar $ud\bar{s}\bar{s}$ state with $J^P=1^-$
was suggested 
in analogy with the $\Theta^+$ by Burns et al.\cite{Close-4q},
and its mass was predicted to be $\sim$ 1.6 GeV. 
The isoscalar tetraquark $ud\bar{s}\bar{s}$ in the flavor $\bar{10}$ group 
was called $\vartheta^+$-meson in Ref.[\ref{Close-4q}] in the association with 
the $\Theta$ baryon.
In this paper, we investigate the $\vartheta^+$-meson($J^P=1^+$)
with a constituent quark model.
The theoretical method of the calculations is the same as that applied
to the study of pentaquark and tetraquark in 
Refs.[\ref{ENYO-penta},\ref{ENYO-tetra}].
Namely, we apply the flux-tube quark model with antisymmetrized molecular
dynamics(AMD)\cite{ENYObc} to $4q$ systems.
Based on the picture of a flux-tube model, we ignore the coupling 
between the configurations $(qq)_{\bar{3}}(\bar{q}\bar{q})_3$
and $(q\bar{q})_{1}(q\bar{q})_{1}$, and solve 
the $4q$ dynamics with the variational method only 
in the model space within the exotic color configuration
$(qq)_{\bar{3}}(\bar{q}\bar{q})_3$. By assuming that 
the $f_1$ meson in the 1.4$\sim$1.5 GeV mass region as the $4q$ state,
we predict the mass and the width of the tetraquark $\vartheta^+$($J^P=1^+$).


The method of AMD is a variational method.
The adopted Hamiltonian is the same as that of previous 
works\cite{ENYO-penta,ENYO-tetra}.
The Coulomb and color-magnetic terms of the OGE potential 
and the string potential are taken into account.
The parameters in the Hamiltonian are chosen to reasonably 
reproduce the normal hadron spectra\cite{ENYO-tetra}.
In order to evaluate the $\vartheta^+(1^+)$ mass and the width, 
we adopt the observed data of the $f_1$-mesons($f_1(1420)$ and $f_1(1510)$) 
in the 1.4$\sim$1.5 GeV region as an input.
The details of the formulations for fourquark systems are explained in
Ref.[\ref{ENYO-tetra}].


As mentioned in \cite{ENYO-tetra}, in order to predict the 
$\vartheta$-meson mass, we need to determine the  
mass shift parameter $M_0$ in the string potential for  
fourquark $(qq)_{\bar{3}}(\bar{q}\bar{q})_{3}$ systems.
Here, we use the $f_1$-meson mass as the input to 
determine the mass shift. 
In the mass region 1$\sim$1.6 GeV, three $f_1$-mesons,
$f_1(1285)$, $f_1(1420)$, and $f_1(1510)$ are known,
though the $f_1(1510)$ is not well established \cite{PDG}.
In the $P$-wave $q\bar{q}$ state, two $f_1$-mesons are expected to 
appear in this energy region as the partners of the $q\bar{q}$ nonet. 
It is considered that the lower one is dominated by 
the light-quark $u\bar{u}$ and $d\bar{d}$ components($n\bar{n}$), 
and the major component of the higher one is the $s\bar{s}$ state.
In the general interpretation, the lowest $f_1$-meson($f_1(1285)$) is 
regarded as the $n\bar{n}$ state.
However, there are two candidates 
($f_1(1420)$ and $f_1(1510)$) for the $q\bar{q}$ partner of the $f_1(1285)$, 
and the assignment is not confirmed yet. 
In the constituent quark model calculation \cite{Isgur-2q},
the masses of the two $1^{++}$ $q\bar{q}$ states in the $P$-wave 
$q\bar{q}$ nonet are 1.24 and 1.48 GeV. 
The theoretical mass spectra of the $1^{++}$ $q\bar{q}$ states 
seems to be consistent with the experimental ones
if $f_1(1510)$ is assigned to be the partner of the $f_1(1285)$ in the 
flavor nonet.
This is consistent with the assignment in Ref.[\ref{Godfrey}].
On the other hand, an alternative interpretation of the 
$f_1(1285)$ and $f_1(1420)$ being the $q\bar{q}$
partners is claimed in Refs.[\ref{PDG},\ref{close97},\ref{close02}].
These interpretations lead to an indication that one of the 
$f_1(1420)$ or $f_1(1510)$ may be a non-$q\bar{q}$ meson while the other 
and $f_1(1285)$ can be understood as the partners of 
the conventional $P$-wave $q\bar{q}$ states.

By ignoring the $q\bar{q}$ annihilation, we calculate the mass of the 
$J^{PC}$=$1^{++}$ $(us)_{\bar{3}}(\bar{u}\bar{s})_3$ state for 
the $f_1$-meson with the present framework 
in the same way as for the tetraquark $\vartheta$-meson.
We adjust the mass shift parameter $M_0$ 
by fitting the mass of the $f_1(1420)$ or the $f_1(1510)$.
Then we get the $\vartheta^+(1^+)$ mass around 1.4 GeV.

The obtained $\vartheta^+(1^+)$ wave function is dominated by 
the component with the spin-zero $(ud)_{\bar{3}}$ and the spin-one
$(\bar{s}\bar{s})_3$ in the spatially symmetric configuration, $(0s)^4$.
This is consistent with the naive expectation in the diquark picture
because the spin-zero $(ud)_{\bar{3}}$ gain the color-magnetic
interaction while only the spin-one
configuration is allowed in the spatially symmetric 
$(\bar{s}\bar{s})_3$ pair. As a result of the energy gain of the 
color magnetic interaction, the $\vartheta^+(1^+)$ mass 
is slightly lower than the fourquark $f_1$ mass.

Next we discuss the width of the $\vartheta^+(1^+)$ meson.
As mentioned above, we suggest that the $\vartheta^+(1^{+})$-meson
may appear in the energy region $\sim$1.4 GeV near the $KK^*$ threshold.
The expected decay modes are $KK^*$ and $KK\pi$.
If the branching into $KK^*$ is small enough, the width should be
narrow because the phase space for the three-body decays is suppressed
in general. In order to discuss the stability of the $\vartheta^+$-meson,
we consider only the two-hadron decay and
give a rough estimation of the $\vartheta^+$ width
assuming that the coupling of the fourquark $f_1$
with the $K\bar{K}^*$ and $c.c.$ is the same as 
that of the $\vartheta^+$ with the $KK^*$.
By considering the phase space for the two-body decays, 
we predict the width of $\vartheta^+(1^{+})$ to be 
$\Gamma_{\vartheta}=20 \sim 50$ MeV.


In summary, we discussed the possibility of the $J^P=1^+$ state of the
isoscalar tetraquark(S=+2), $\vartheta^+$-meson, 
with the $ud\bar{s}\bar{s}$ content.
If a $f_1$ meson in the mass region $1.4-1.5$ GeV consists of 
four quarks $ns\bar{n}\bar{s}$,
the mass of the isoscalar $ud\bar{s}\bar{s}$($\vartheta^+$-meson) 
state with $J^P=1^+$ is expected to be lower than 
that of the $f_1$ meson.
Within a flux-tube quark model, 
a possible resonant state of 
$ud\bar{s}\bar{s}(J^{P}=1^{+})$ is suggested to appear at $\sim$ 1.4 GeV 
with the width ${\cal{O}}(20\sim 50)$ MeV.
We propose that the $\vartheta^+$-meson is the good candidate for the 
tetraquark search, which would be observed in the $K^+K^+\pi^-$ decay channel.

Recently, the $\theta^+$($1^-$) and the $\theta^+$($0^+$)
were predicted by Burns et al.\cite{Close-4q}
and  by Karlier and Lipkin \cite{karliner04}, respectively.
It should be pointed out that 
the allowed decay channels are different among these
three predictions $J^P=1^+$, $1^-$, and $0^+$ states of 
the $\vartheta^+$-mesons. 
The $K^+K^+\pi^-$ decay from 
the $\vartheta$($J^{P}=1^{+}$) predicted in the present work
is suitable for the experimental tetraquark search.

As for the other candidates of the $4q$ states,
it has been theoretically suggested that the 
scalar mesons like $f_0(600)$, $f_0(980)$ and $a_0(980)$ 
below 1 GeV might be interpreted as $4q$ states.
We calculated the corresponding fourquarks  
with the present framework. Then, we found that the 
masses of fourquarks with the exotic color configurations
$(qq)_{\bar{3}}(\bar{q}\bar{q})_{3}$ are  much higher 
than these light scalar mesons. It indicates that these
scalar mesons might be other than  
the dominant $(qq)_{\bar{3}}(\bar{q}\bar{q})_{3}$ state, but might be
the hybrids of $P$-wave $q\bar{q}$ and fourquark components
with meson-meson tails in the outer region as argued in Ref.[\ref{close02}].

The authors would like to thank to H. Nemura and H. Hidaka 
for the valuable discussions.
This work was supported by Japan Society for the Promotion of 
Science and Grants-in-Aid for Scientific Research of the Japan
Ministry of Education, Science Sports, Culture, and Technology.
The calculations of this work is supported by the computer system in YITP.

\end{document}